\documentclass[osajnl,twocolumn,showpacs,superscriptaddress,10pt]{revtex4-1}

\usepackage{graphicx}
\usepackage{dcolumn}
\usepackage{bm}
\usepackage{amsmath,amssymb,graphicx}

\begin{document}

\title{
Observation of a classical cheshire cat in an optical interferometer
}

\author{David P. Atherton}
\affiliation{Department of Physics, University of Nevada, Reno NV 89557, USA}

\author{Gambhir Ranjit}
\affiliation{Department of Physics, University of Nevada, Reno NV 89557, USA}

\author{Andrew A. Geraci}
\email{ageraci@unr.edu}
\affiliation{Department of Physics, University of Nevada, Reno NV 89557, USA}

\author{Jonathan D. Weinstein}
\email{weinstein@physics.unr.edu}
\homepage{http://www.physics.unr.edu/xap/}
\affiliation{Department of Physics, University of Nevada, Reno NV 89557, USA}


\begin{abstract}

A recent neutron interferometry experiment claims to demonstrate a paradoxical phenomenon dubbed the ``quantum Cheshire Cat'' \cite{Denkmayr2014}. We have reproduced and extended these results with an equivalent optical interferometer. The results suggest that the photon travels through one arm of the interferometer, while its polarization travels through the other. However, we show that these experimental results belong to the domain where quantum and classical wave theories coincide; there is nothing uniquely quantum about the illusion of this cheshire cat.

\end{abstract}

\maketitle


The concept of quantum weak measurements was introduced over two decades ago \cite{PhysRevLett.60.1351} and experimentally realized soon after \cite{PhysRevLett.66.1107}.
Since that time, weak measurements have found utility in a variety of experiments \cite{RevModPhys.86.307}, such as the amplification of small signals \cite{Hosten08022008}, quantum feedback to prepare and stabilize desired quantum states \cite{Haroche2011}, the development of new methods to directly measure quantum states \cite{DirectWavefunctionMeasurement2011}, and as a probe of the nonclassical features of quantum mechanics \cite{1367-2630-9-8-287, PhysRevLett.102.020404}.

However,  there has long been controversy over the interpretation of weak measurements \cite{PhysRevLett.62.2325, PhysRevD.40.2112}, including work showing that  weak measurement experiments can be interpreted as classical phenomena
\cite{1367-2630-15-7-073022}.
This controversy has been renewed with recent work on the ``quantum Cheshire Cat''.
The concept of a quantum Cheshire Cat was introduced  by Aharonov \textit{et. al.}   \cite{Aharonov2013}. The proposed experiment considered weak measurements of photons in an interferometer. For an appropriately pre- and post-selected ensemble, a weak measurement of photon polarization produces a nonzero value in only one arm, while a weak measurement of the photon itself produces a nonzero value in the other arm.

Soon after, Denkmayr \textit{et. al.} reported the observation of a quantum Cheshire Cat in a neutron interferometry experiment \cite{Denkmayr2014}. This experiment used a different technique than proposed by Aharanov \textit{et. al.}: the neutron was measured with an absorber and its spin measured through a unitary rotation with a magnetic field. 
In this experiment, it was found that for appropriate pre- and post-selection, the post-selected signal is significantly affected  by neutron measurement in only one arm, and spin measurement in only the other.

Because the photon's polarization is isomorphic to the neutron's spin of 1/2, this experiment has a photonic 
equivalent.
Using an optical interferometer, we reproduce the experimental results of the neutron experiment and extend the work to simultaneous weak measurements of  position and  polarization.

\begin{figure}[h]
    \begin{center}
    \includegraphics[width=\linewidth]{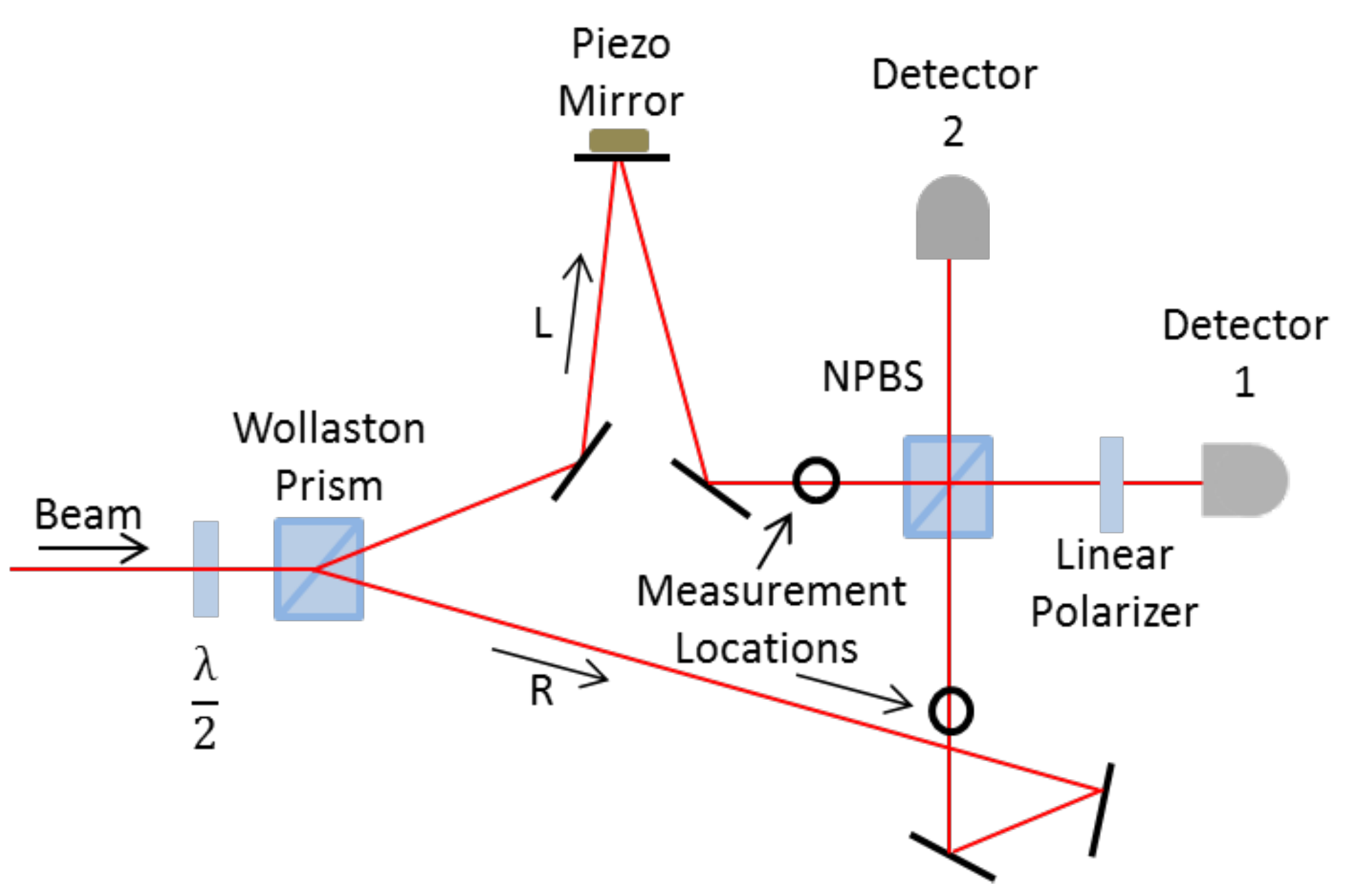}
    \caption{ (Color online)
    Experimental setup, as described in the text. Absorbers and waveplates are placed at the two ``measurement locations''.
    %
        \label{fig:setup}}
    \end{center}
\end{figure}

The experimental setup is illustrated in Fig. \ref{fig:setup}. We obtain 780~nm photons from a  fiber-coupled diode laser stabilized to an atomic rubidium transition.
Typical powers used were 0.8 mW.
A Wollaston prism  splits this light into two beams: $L$ and $R$, with orthogonal linear polarizations $H$ and $V$ respectively.
A half-wave plate before the prism is used to roughly balance the power of the two beams.
Thus, the photons are prepared in the preselected state
$$ |\Psi \rangle = \frac{1}{\sqrt{2}} \Big( |H\rangle |L \rangle + | V\rangle | R \rangle \Big) $$
%
%

After propagating along paths $L$ and $R$, the beams are recombined on a non-polarizing cube beamsplitter (NPBS).
The  interferometer outputs are measured with two silicon photodetectors.
An absorptive linear polarizer, which passes polarization $H$, is placed in front of detector 1.
Thus, detector 1 post-selects the state
$$ |\Psi \rangle = \frac{1}{\sqrt{2}} | H \rangle \Big( |L \rangle  + |R \rangle \Big) $$
One of the mirrors in path $L$ is mounted on a piezoelectric actuator to allow modulation of the length of path $L$ and thus the relative phase of the two beams.
This is modulated with a sufficient amplitude to obtain a phase change in excess of $2\pi$.

 Absorbers and polarization rotators (half-wave plates) can be placed in paths $L$ and $R$ immediately before the non-polarizing beamsplitter to perform position and polarization ``measurements''.


Position measurements were obtained by placing an absorptive neutral density filter in either path $L$ or $R$.  A drop in power on the detector implies that the filter absorbed some of the photons from that path.  If there is no drop in power, one can conclude that --- for the  pre- and post-selected states --- no photons traveled through that path.

\begin{figure}[h]
    \begin{center}
    \includegraphics[width=\linewidth]{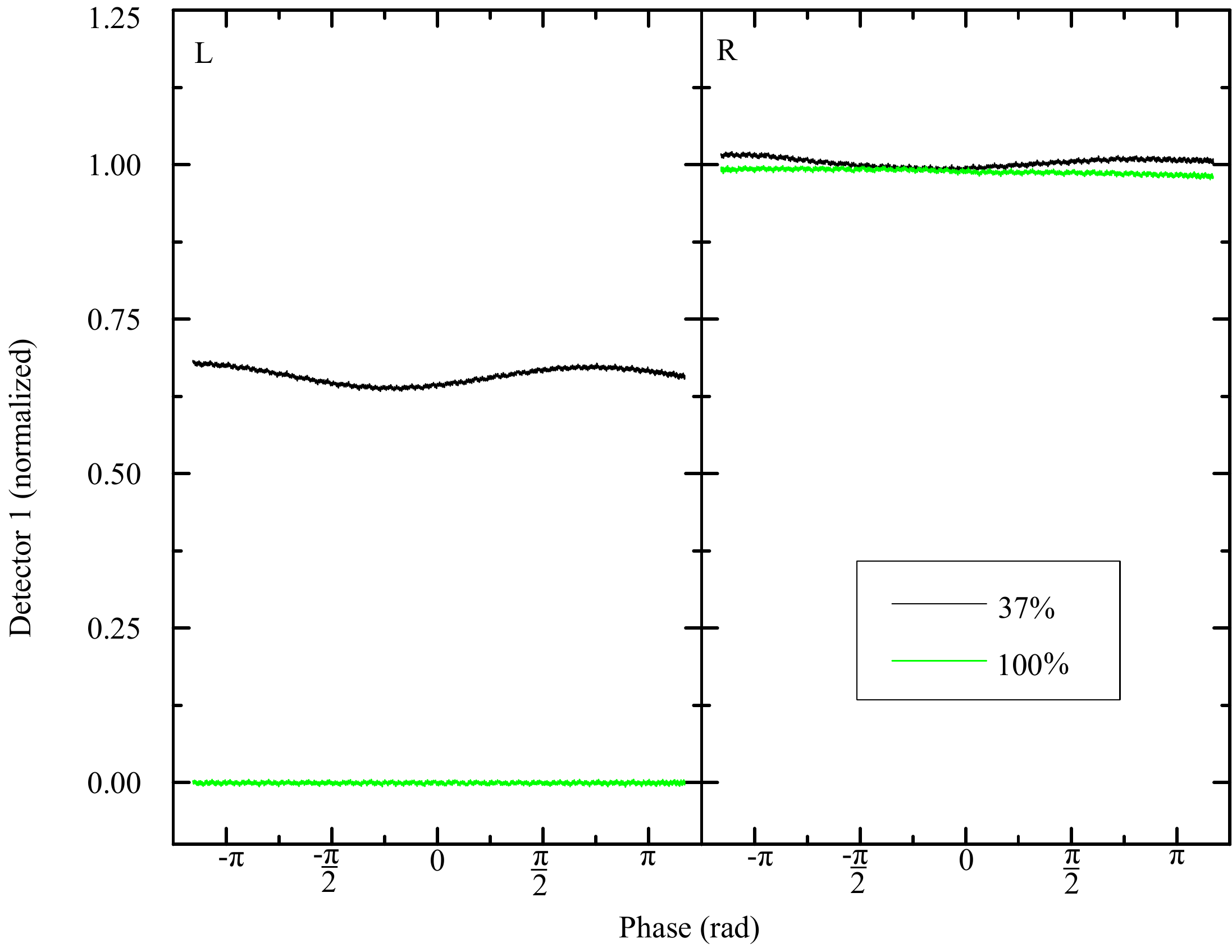}
    \caption{ (Color online)
    Photon position measurement:
    power as measured by photodetector 1 as a function of the relative phase between the optical paths. Shown are data for 37\% and 100\% absorbers placed in either path $L$, as shown in the left panel, or path $R$, as shown in the right panel.
    All signals are normalized to the level of an empty spectrometer.
    Because the interferometer is not actively locked to the laser, the phase drifts between measurements.
    %
        \label{fig:location}}
    \end{center}
\end{figure}

The data for position measurement is shown in Fig. \ref{fig:location}, showing the experimental results for absorbers in either path $L$ or $R$.
Both weak measurement and strong measurement cases are shown.


We would expect this signal to have no dependence on the relative phase delay of the two paths; the modulation we see is due to imperfections in our polarization.
A drop in power was observed when the filter was placed in path $L$.  However, no significant change was observed when the absorber was placed in path $R$.
This indicates that, for the pre- and post-selected states and no polarization rotation, the photons traveled through path $L$.

\begin{figure}[h]
    \begin{center}
    \includegraphics[width=\linewidth]{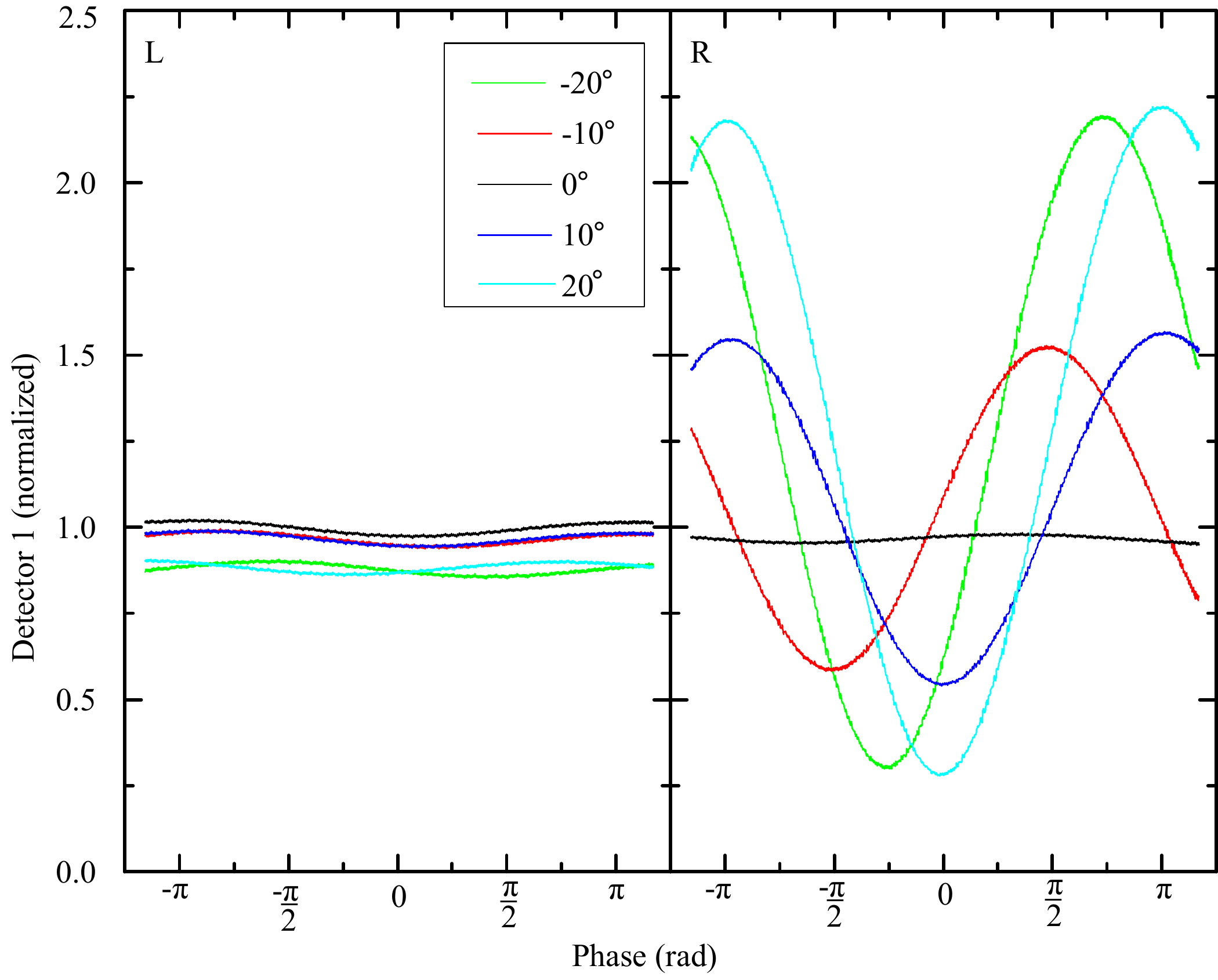}
    \caption{ (Color online)
    Photon polarization measurement. The data is similar to Fig. \ref{fig:location}, except a polarization rotator is placed in either path $L$, as shown in the left panel, or path $R$, as shown in the right. Polarization rotation angles as labeled in the figure legend.
        \label{fig:polarization}}
    \end{center}
\end{figure}

To measure polarization via rotation (equivalent to the measurement-of-spin-via-rotation of Denkmayer \textit{et. al.}), we place a half-wave plate in either path $L$ or $R$. 
The resulting data is shown in Fig. \ref{fig:polarization}.
If the polarization in path $L$ is rotated, there is little effect other than a small drop in power which is second-order in the rotation angle.
If the polarization in path $R$ is rotated, there is a strong effect. For the pre- and post-selected states, this suggests the photon's polarization has traveled through arm $R$.


\begin{figure}[h]
    \begin{center}
    \includegraphics[width=\linewidth]{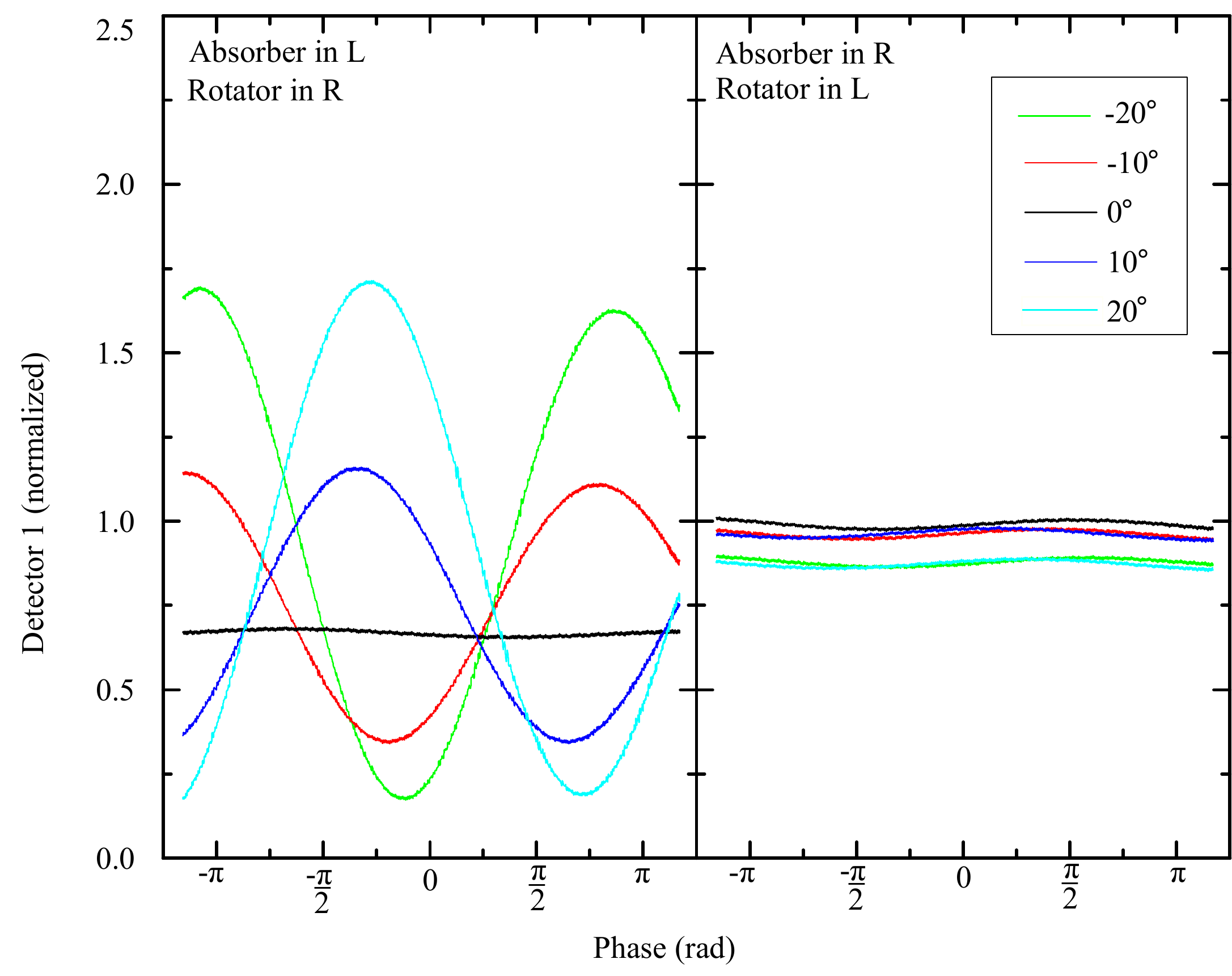}
    \caption{ (Color online)
    Simultaneous photon position and polarization measurement.
    The data is similar to Fig. \ref{fig:location}, except in the left panel a polarization rotator is placed in  path $L$ and a 37\% absorber is placed in path $R$, and in the right panel  a polarization rotator is placed in  path $R$ and a 37\% absorber is placed in path $L$. 
        \label{fig:simultaneous}}
    \end{center}
\end{figure}

To measure position and polarization simultaneously, we place a weak absorber in one arm and apply a small rotation in the other, as shown in Fig. \ref{fig:simultaneous}.
When rotations are performed in path $L$ along with simultaneous absorption in path $R$ we see no significant changes in the data.  However, when  rotations are performed in path $R$ and  absorption in path $L$ we see a significant effects to both absorption and polarization. This suggests that, for the pre- and post-selected states, the photon travels through $L$ while its polarization simultaneously travels through $R$.

\emph{Interpretation:} 
We see that the post-selected output of the interferometer is sensitive to absorption in only one arm, and weak rotations in only the other.
The ``weak-measurement'' interpretation of these results is identical to that of
Denkmayr \textit{et. al.} \cite{Denkmayr2014}, and we do not reproduce it here.
Instead, we analyze our system with classical waves.

We begin with a polarized wave $E_0 \cos (\omega t) \hat{h}$ in the left arm and $E_0 \cos (\omega t) \hat{v}$ in the right, where $\hat{h}$ and $\hat{v}$ are orthogonal linear polarizations.


We then introduce a phase shift $\phi$ in the left arm; attenuators with transmissions $T_{L,R}$ in the left and right arms, respectively; and rotations of the polarization $\theta_{L,R}$ in the left and right arms, respectively. The beams are combined on a beamsplitter and passed through a linear polarizer which passes the $\hat{h}$ polarization.

On the detector, we then have an electric field
%
$$\frac{E_0}{2} \left ( \sqrt{T_L}\cos\theta_L \cos(\omega t + \phi) + \sqrt{T_R}\sin\theta_R \cos(\omega t) \right) \hat{h}$$
%
%
%
This gives a time-averaged intensity
\begin{eqnarray*}
I &\propto&   T_L \cos^2 \theta_L + T_R \sin^2 \theta_R \\
&& + 2\cos(\phi) \sqrt{T_L T_R} \cos\theta_L  \sin\theta_R
\end{eqnarray*}
Unsurprisingly, this reproduces the observed results, as cataloged below.

First, in the absence of rotation, $I \propto   T_L $. Thus, the signal depends only on absorption in the left arm. This is true for weak or strong absorption.

Second, in the absence of absorption, $I \propto   \cos^2 \theta_L + \sin^2 \theta_R + 2 \cos(\phi) \cos\theta_L  \sin \theta_R$. For the case of infinitesimal rotations, we expand this to first order in $\theta_{L,R}$ to find
$I \propto 1 + 2 \theta_R \cos(\phi) $. Thus, the signal is only affected by small polarization rotations in the right arm.

Finally, in the presence of simultaneous infinitesimal absorption and infinitesimal rotation,  the DC level is affected only by $T_L$, and the interference fringes are affected only by $\theta_R$.

The classical model reproduces the observed measurements qualitatively. There are  quantitative differences on the level of 10 to 20\%, which stem from imperfections in our polarization optics, beam overlap, and slight power imbalances between the two arms.



\emph{Conclusion:}
While the ``quantum cheshire cat'' may 
appear to reflect new physics, all that has been demonstrated by our experimental results --- as well as the results of Denkmayr \textit{et. al.} \cite{Denkmayr2014} --- is the following: in this situation,  ensemble averages of unentangled quantum mechanical particles give results consistent with the predictions of classical wave interference.
This stands in  contrast to uniquely quantum phenomena such as sub-poissonian light \cite{PhysRevLett.39.691}, interferometry with entangled particles \cite{Leibfried04062004},  and squeezed light interferometry \cite{PhysRevLett.59.278}, all of which yield uniquely quantum, nonclassical behavior.
%
In cheshire cat experiments performed to date, there is no more separation of photon from its polarization (nor neutron from its spin) than there is separation of amplitude from polarization for classical waves. The only oddity is the age-old quantum mechanics mystery of particles whose ensemble averages behave as waves.

\emph{Note:}
While preparing this manuscript, we became aware of the work of Correa \textit{et. al.}, which theoretically analyzes the cheshire cat, and shows both the neutron experiment of Denkmayr \textit{et. al.} \cite{Denkmayr2014} and the proposed experiments of  Aharonov \textit{et. al.}   \cite{Aharonov2013} can be interpreted as simple quantum interference
 \cite{arXiv:1409.0808}.


\emph{Acknowledgements:} This material is based upon work supported by the National Science Foundation under Grant No. PHY 1265905 and Grant No. PHY 1205994.

\end{document}